\documentstyle[12pt]{article}                                 
\usepackage{graphicx}

\begin{document}                                             

%%%%%%%%%%%%%%%%%%%%%%%%%%%%%%%%%%%%%%%%%%%%%%%%%%%%%%%%%%%%%%%%%%%%%%%%
\centerline{\Large \bf How does breakup influence near-barrier fusion }

\centerline{\Large \bf of weakly bound light nuclei ? }
%%%%%%%%%%%%%%%%%%%%%%%%%%%%%%%%%%%%%%%%%%%%%%%%%%%%%%%%%%%%%%%%%%%%%%%%

\vspace{9ex}

\leftline{\Large C. Beck, N. Rowley, M. Rousseau, F. Haas, P. Bednarczyk,} 
\leftline{\Large S. Courtin, N. Kintz, F. Hoellinger, P. Papka, S. Szilner,}
\leftline{\Large A. S\`anchez i Zafra, A. Hachem, E. Martin, and O. Stezowski} 

\vspace{2ex} 

\leftline{\it Institut de Recherches Subatomiques, UMR7500, CNRS-IN2P3 et}
\leftline{\it Universit\'{e} Louis Pasteur Strasbourg} 
\leftline{\it 23, rue de Loess, B.P. 28}
\leftline{\it F-67037 Strasbourg Cedex 2, France }

\vspace{3ex}

\leftline{\Large A. Diaz-Torres}

\vspace{2ex}

\leftline{\it Institut f\"{u}r Theoretische Physik, Goethe-Universit\"at
Frankfurt,} 
\leftline{\it Robert Mayer Strasse 8-10,}
\leftline{\it D-60054 Frankfurt am Main, Germany}

\vspace{3ex}

\leftline{\Large  F.~A. Souza, A. Szanto de Toledo, A. Aissaoui, N. Carlin}
\leftline{\Large  R. Liguori Neto, M.~G. Munhoz, J. Takahashi,}
\leftline{\Large  A.~A.~P. Suade, M.~M. de Moura, and E.~M. Szanto}

\vspace{2ex}

\leftline{\it Laborat\'orio Pelletron, Departamento de Fisica Nuclear,} 
\leftline{\it Instituto de Fisica, Universidade de S\~{a}o Paulo, } 
\leftline{\it CP 66318, 05315-970, S\~{a}o Paulo-SP, Brasil }

\vspace{3ex}

\leftline{\Large K. Hagino}

\vspace{2ex}

\leftline{\it Department of Physics, Tohuku University,} 
\leftline{\it 980-77 Sendai, Japan }

\vspace{3ex}

\leftline{\Large I.~J. Thompson}

\vspace{2ex}

\leftline{\it Physics Department, University of Surrey,}
\leftline{\it GU2 7XH Guildford, United Kingdom}

\newpage

\leftline{\Large \bf ABSTRACT:}

\vspace{4mm}

\noindent
The influence on the fusion process of coupling to collective degrees of
freedom has been explored. The significant enhancement of the fusion cross
section at sub-barrier energies was understood in terms of the dynamical
processes arising from strong couplings to collective inelastic excitations of
the target and projectile. However, in the case of reactions where breakup
becomes an important process, conflicting model predictions and experimental
results have been reported in the literature. Excitation functions for sub- and
near-barrier total (complete + incomplete) fusion cross sections have been
measured for the $^{6,7}$Li+$^{59}$Co at the {\sc Vivitron} facility and at the
8UD {\sc Pelletron} tandem facility using standard $\gamma$-ray techniques. The
data extend to medium-mass systems previous works exploring the coupling
effects in fusion reactions of both lighter and heavier systems. Results of
Continuum-Discretized Coupled-Channel (CDCC) calculations indicate a small
enhancement of total fusion for the more weakly bound $^{6}$Li at sub-barrier
energies, with similar cross sections for both reactions at and above the
barrier. A systematic study of $^{4,6}$He induced fusion reactions with the
CDCC method is in progress. The understanding of the reaction dynamics
involving couplings to the breakup channels requires the explicit measurement
of precise elastic scattering data as well as yields leading to the breakup
itself. Recent coincidence experiments for $^{6,7}$Li+$^{59}$Co are addressing
this issue. The particle identification of the breakup products have been
achieved by measuring the three-body final-state correlations. 

\vspace{4mm}

\noindent
{\bf PACS} number(s): 25.70.Jj, 25.70.Mn, 25.70.Gh, 24.10.Eq

\newpage

\section{Introduction}

\medskip

The study of fusion reactions in the vicinity of the Coulomb barrier provides a
fascinating challenge for theories of quantum tunneling leading to an
irreversible complete fusion of the interacting nuclei into the compound
nucleus (CN) \cite{Balantekin98,Khalili03,Thompson04,Dasgupta98}. A great
experimental effort involving both (loosely bound) stable and unstable nuclei
has been devoted to investigate the specific role of the breakup channel
\cite{Dasgupta98}. The recent availability of light-mass radioactive ion beams
such as $^{6}$He \cite{Kolata98,Trotta00,Raabe04}, $^{11}$Be \cite{Yoshida96},
and $^{17}$F \cite{Rehm98}, motivated the investigation of fusion reactions
involving very weakly bound nuclei around and below the Coulomb barrier.\\ 

\noindent
The fusion probability is sensitive to the internal structure of the
interacting ions as well as to the influence of the other competiting
mechanisms such as nucleon transfer and/or breakup which are known to affect the
fusion features \cite{Dasgupta98}. The fusion cross section enhancement
generally observed at sub-barrier energies is understood in terms of dynamical
processes arising from couplings to collective inelastic excitations of the
target and/or projectile. However, in the case of reactions where at least one
of the colliding ions has a sufficient low binding energy so that breakup becomes an
important process, conflicting experimental
\cite{Kolata98,Trotta00,Raabe04,Yoshida96,Rehm98,Takahashi97,Dasgupta99,Tripathi02,Anjos02,Dasgupta02,Wu03,Bychowski04}
and theoretical results are reported
\cite{Takigawa91,Hussein92,Dasso94,Canto95,Yabana95,Hagino00,Mohr00,Diaz02a,Diaz02b,Diaz03,Kim02,Alamanos02,Keeley02,Cardenas03,Rusek04,Mackintosh04}.
\\ 

\noindent
Clearly a full understanding of the effects of breakup on near-barrier fusion
requires systematic and detailed measurements covering a wide range of systems
and energies. Here we choose to study both the total fusion
\cite{Beck03,Souza03,Szanto03} and breakup \cite{Szanto03,Szanto04,Souza04} of
$^{6,\,7}$Li with the intermediate-mass target $^{59}$Co. Fusion measurements
have been performed by detecting characteristic $\gamma$ rays emitted from the
resulting evaporation residues \cite{Beck03}, which could in principle also
allow us to distinguish between the different kind of fusion processes:
complete fusion (CF) and incomplete fusion (ICF). CF requires the possibility
of fusion through the CN formation containing all the nucleons of both the
intact projectile and the target. If only part of the projectile-like fragments
may emerge from the interaction region with a compound system being formed then
ICF is defined (in this case the breakup process is followed by fusion
\cite{Dasgupta99,Hussein92,Hagino00,Diaz03,Beck03}). Breakup yield measurements have
been achieved by measuring the three-body final-state correlations with charged
particle techniques \cite{Szanto03,Szanto04,Souza04}. \\ 

\noindent
The present work extend the fusion study for medium-mass systems by
exploring the coupling effects (hindrance versus enhancement) in the
framework of the Continuum-Discretized Coupled-Channel {\sc Cdcc}
method \cite{Diaz03}. Whereas a small enhancement of total (a sum of CF
and ICF cross sections) fusion is observed for the more weakly bound $^{6}$Li
at sub-barrier energies, this enhancement is predicted to be much more
significant for halo nuclei such as $^{6}$He. The effect disappears above the
barrier \cite{Beck03}. \\

\newpage
In this work we present the fusion data for $^{6,7}$Li+$^{59}$Co in Sec. II
analysed by detailed {\sc Cdcc} calculations in Sec. III. Their elastic data as
well as the corresponding exclusive light charged particle experiment (which
analysis is still in progress) is described in Sec. IV while a summary with
very preliminary conclusions and some perspectives using the {\sc Cdcc}
approach are given in Sec. V. 

\newpage

\section{Experimental $\gamma$-ray method and data analysis}

\medskip

The $^{6}$Li+$^{59}$Co and $^{7}$Li+$^{59}$Co fusion reactions are used to
investigate the effect of breakup on the fusion cross section
\cite{Beck03,Souza03,Szanto03,Szanto04,Souza04}. These fusion-evaporation
measurements help to establish the influence of the projectile breakup on the
fusion process at near-barrier energies and show how the mass of the target
affects the process, as well as the ICF yield. Experiments have been performed
either at the {\sc Vivitron} electrostatic tandem accelerator of the IReS
Strasbourg~\cite{Beck03} or at the 8UD {\sc Pelletron} tandem facility of the
University of S\~ao Paulo~\cite{Souza03,Szanto03}. Standard $\gamma$-ray
techniques have been used for the Strasbourg measurements: the $\gamma$-ray
events were detected with part of the {\sc Garel+} spectrometer
array~\cite{Beck03} configured with 14 Compton-suppressed, high-efficiency {\sc
Eurogam}-type Hp Ge detectors together with one LEPS (Low-energy Photon
Spectrometer) detector. The absolute efficiency (for a calibrated $^{60}$Co
source) for {\sc Garel+} was 1.2$\pm$0.2\% and 0.48$\pm$0.04\% for the S\~{a}o
Paulo Ge detector setup. More details of the experimental setup and the
analysis procedures are given in Ref.~\cite{Beck03}. \\ 

\noindent
It is important to notice that the complete data set presented in this work for
the $^{6}$Li+$^{59}$Co and $^{7}$Li+$^{59}$Co fusion reactions using the
$\gamma$-ray spectroscopy method is in very good agreement with
$^{6,7}$Li+$^{64}$Zn data obtained with charged particle techniques
\cite{Gomes04}. Unfortunately, it has not been possible to clearly resolve the
ICF and CF components for the $^{59}$Co target results~\cite{Beck03}. In
principle, it should be possible to estimate the ICF contribution by studying
the specific population pattern of states in the ER's in the context of
statistical-model calculations for these population patterns. Although the
current data \cite{Beck03} do not have high-statistics to perform a definite
analysis, Signorini {\it et al.} \cite{Glodariu03} have been able to perform
statistical-model calculations (with the PACE2 code). These imply that $\alpha$
({\it t}) ICF components are rather weak. In the following the ER cross
sections will be considered to be the total fusion cross sections when compared
with the various theoretical calculations
\cite{Hagino00,Diaz02a,Diaz03,Beck03}. \\ 

\noindent
Fig.~1.(a) displays the experimental excitation functions of the total fusion
cross sections measured for the two studied reactions. The average fusion
excitations functions can be obtained by fitting the data of Fig.~1.(a) 
can be obtained with the phenomenological one-dimensional barrier penetration
model (SBPM) of Wong \cite{Wong73}. The use of this crude parametrization is
still justified in the energy region near the Coulomb barrier when inelastic
and transfer channel coupling do not affect the total fusion cross section
significantly. The results of such SBPM fits are compared for the quantity 
$\sigma_{fus}$E$_{c.m.}$ plotted as a function of the center of mass energy
$E_{c.m.}$ in Fig.~1.(b) for the two reactions $^{6}$Li+$^{59}$Co (in blue
color) and $^{7}$Li+$^{59}$Co (in red color). The relevant parameters are the
barrier radius {\it R}, the ``curvature" $\hbar$$\omega$ of the barrier, and
the barrier height V$_{B}$. For near-barrier energies the three parameters
control the position and slope of the curves plotted in Fig.~1.(b). The values
of the parameters are found to be consistent with the current systematics
\cite{Anjos02,Gomes04} with the noticeable exception of the $\hbar$$\omega$
value being too high (8.1 MeV) for $^{6}$Li. Although SBPM fitting can still be
done even if inelastic and/or transfers are important, this anomalous value may
give some indication that the direct breakup/scattering process might have a
significant influence on the total fusion cross section.\\ 

\noindent
In order to better isolate the effects of possible couplings, it is important
to use a clear reference when an enhancement and/or a suppression is defined.
Therefore the ratio R = $\sigma$($^{6}$Li)/$\sigma$($^{7}$Li) between the total
fusion cross sections for $^{6}$Li and $^{7}$Li induced reactions shown in
Fig.~1.(a) is calculated for this purpose. It should be noted that the $^{7}$Li
induced reaction is used as reference due to its higher binding energy (2.47
MeV) when compared to 1.475 MeV for $^{6}$Li. This ratio is displayed in Fig.~2
as a function of E$_{c.m.}$ and compared to CC calculations
\cite{Hagino00,Diaz02a} discussed in the next Section. \\ 

\newpage

\section{Coupled-Channels calculations}

\medskip

Since the coupling of the relative motion between colliding nuclei to the
inelastic (and transfer) channels is known to enhance the fusion cross section
at sub-barrier energies~\cite{Dasso94}, CC effects are usually taken into
account in the theoretical descriptions of the fusion process. The question of
whether the breakup process may strongly influences the fusion processes (i.e.
as a sub-barrier enhancement) remains open. Therefore we have chosen to apply
two different CC approaches \cite{Hagino00,Diaz02a} to the $^{6}$Li+$^{59}$Co
and $^{7}$Li+$^{59}$Co fusion reactions presented in this work. The solid curve
in Fig.~2 shows the ratio calculated the model of Wong \cite{Wong73} using
independent fits of two SBPM parametrizations shown in Fig.~1.(b). If the only
difference between the two lithium isotopes was the $A^{1/3}$ dependence of the
radius, one would expect a simple shift in energy due to the corresponding
difference in barrier heights. This shift should be around 0.14 MeV and the
dotted curve gives the resulting ratio in Fig.~2; the latter clearly goes in a
direction opposite to the experimental data at sub-barrier energies. \\ 

\noindent
The other curves plotted in Fig.~2 correspond to two CC calculations
\cite{Hagino00,Beck03} with and without reorientation effects (microscopic
differences in the structure of the two Li isotopes lead to different
reorientation terms in the channel couplings) performed by using the {\sc
Ccfull} code \cite{Hagino99}. This code solves the Schr\"odinger equation and
the coupled equations exactly, making only the iso-centrifugal approximation.
The fusion cross sections are calculated using an incoming wave boundary
condition and taking a Woods-Saxon form for the nuclear potential
\cite{Broglia91}. The potential parameters were taken to be identical for both
projectiles: the depth V$_{0}$ = 74.0 MeV, the radius parameter r$_{0}$ = 1.05
fm, and the diffuseness parameter $a$ = 0.63 fm. This value is very close to
the predictions using the Woods-Saxon parametrization of the Aky\"uz-Winther
potential \cite{Broglia91} which gives $a$ = 0.62 fm and $a$ = 0.63 fm,
respectively, for the $^{6}$Li and $^{7}$Li induced reactions. From the
inspection of Fig.~2 no suppression is observed at energies above the barrier
(one should point out that the experimental data presented in
Ref.~\cite{Beck03} do not allow us to distinguish between CF and ICF
components) and a sub-barrier enhancement is observed for $^{6}$Li. It is
clear, however, that all the details of the data, particularly at sub-barrier
energies, are not reproduced by the present CC calculations. More realistic CC
calculations, taking into account the interplay between projectile breakup and
fusion in the framework of the {\sc Cdcc} approach \cite{Diaz03}, have been
undertaken and presented in Fig.~1. \\ 

\noindent
A traditional approach to discuss the sub-barrier fusion reaction induced by
weakly bound nuclei is to solve the CC equations by discretizing in energy the
continuum states in the projectile nucleus. Here the CC calculations were
performed by using the so-called Continuum-Discretized Coupled-Channel ({\sc
Cdcc}) method~\cite{Diaz02a}. All details concerning the breakup space (number
of partial waves, maximum continuum energy cuttof ...) for $^{6}$Li to obtain
converged total fusion cross sections have been given elsewhere
Refs.~\cite{Diaz02a,Diaz03} (in particular in Table I of \cite{Diaz03}), and
the {\sc Cdcc} scheme is available in a general CC computer code {\sc Fresco} 
\cite{Thompson88}. We would like to stress that in the chosen calculations the
imaginary parts of the off-diagonal couplings have been neglected, while the
diagonal couplings include imaginary parts. We have used short-range imaginary
fusion potentials for each fragment separately. This is equivalent to the use
of incoming boundary conditions in {\sc Ccfull} calculations and guarantees
that at least one of the fragments of the projectile is captured. \\

\noindent
The measured excitation functions that were reported previously in
Refs.~\cite{Beck03,Souza03,Szanto03,Szanto04,Souza04} are here presented in
Fig.~1 (open squares and open circles) with the comparison with predictions
(labelled 'theory' with full triangles and full squares) of {\sc Cdcc}
calculations for both the $^{7}$Li+$^{59}$Co and $^{6}$Li+$^{59}$Co reactions,
respectively. It is shown that the {\sc Cdcc} calculations of Fig.~3 predict
the same significant enhancement of $^{6}$Li (with smaller $\alpha$-breakup
threshold than for $^{7}$Li) fusion cross section. This is due to the fact that
breakup enhances the total fusion cross section just around the Coulomb
barrier, whereas it hardly affects (an enhancement of less then $\approx$
2$\%$) the fusion at energies well above the barrier, as expected. \\ 

\noindent
The $^{6}$He+$^{59}$Co case is much more complicated since $^{6}$He breaks into
three fragments instead of two, and the {\sc Cdcc} method has not yet been
developed for two-nucleon halo nuclei \cite{Diaz02a}. We have used the same
model as for $^{6}$Li+$^{59}$Co case described above. Hence for the
$^{6}$He+$^{59}$Co reaction we assume a two-body cluster structure of $^{6}$He
= $^{4}$He + 2n. The potential between the $\alpha$ particle and the $^{59}$Co
target appears on Table 1 of Ref.~\cite{Diaz03}. Similarly to our previous work
\cite{Diaz03}, the potentials between the fragments and the $^{59}$Co target
are those obtained with the global Broglia-Winther Woods-Saxon parametrization
given in Ref.~\cite{Broglia91} for the Christensen and Winther potential
\cite{Christensen76} (the numerical values are: V$_{o}$=-16.89 MeV,
r$_{o}$=1.09 fm and a=0.63 fm). For the $\alpha$-2n binding potential (0$^{+}$
g.s) we have used the following Woods-Saxon potential: V$_{o}$= -40.796 MeV,
r$_{o}$=1.896 fm and a=0.3 fm. The g.s. binding potential of the $\alpha$
particle and the dineutron provides a 2s bound state of about -0.975 MeV. The
binding potential of the 2$^{+}$ resonant state has also a Woods-Saxon form
with the following parameters: V$_{o}$=-35.137 MeV, r$_{o}$=1.896 fm, a=0.3 fm.
With this potential the energy of the 2$^{+}$ resonant state is 0.83MeV and its
width is 0.075 MeV. To obtain converged (within a 5$\%$) total fusion cross
section we have included: (i) partial waves for $\alpha$-2n relative motion 
up to f-waves ($l$ = 3), (ii) the $^{6}$He fragment-target potential multipoles
up to the octupole term, and (iii) the maximum of the continuum energy is
8 MeV. All resonant and non-resonant continuum couplings including
continuum-continuum couplings were included in the calculation.\\ 

\noindent
Preliminary {\sc Cdcc} calculations using the set of parameters given for the
$^{6}$He+ $^{59}$Co reaction in the previous paragraph are displayed in Figs.~4
and 5. The present calculations do not include neither target excitations nor
transfer channnels. However with crude estimations as those performed for the
$^{6}$Li+$^{59}$Co reaction~\cite{Diaz03} the effect is found to be very small.
In Fig.~4 we compare the total fusion excitation functions of the two
$^{6}$He+$^{59}$Co ({\sc Cdcc} calculations) and $^{6}$Li+$^{59}$Co
(experimental data of Ref.~\cite{Beck03}) reactions. For the $^{6}$He reaction,
the incident energy is also normalized with the Coulomb barrier V$_{B}$ of the
bare potential. The first calculation (dashed line) only include the
reorientation couplings in fusion without breakup. All contiunum and
reorientation couplings are included in fusion with breakup (solid curve). We
can observe that both calculated curves (with and without breakup) give much
larger total fusion cross section for $^{6}$He as compared to $^{6}$Li. We can
also observe that the inclusion of the couplings to the breakup channels
notably increases the total fusion cross section for the whole energy range.
The same conclusions are reached when $^{6}$He+$^{59}$Co ({\sc Cdcc}
calculations) is compared to $^{4}$He+$^{59}$Co (here the {\sc Cdcc}
calculations are fitting the data of Ref.~\cite{McMahan80} remarkably well) in
Fig.~5. \\ 

\noindent

In contrast to stable weakly bound nuclei such as $^{7}$Li and $^{6}$Li,
the questions in the theory of a halo system such as $^{6}$He, its breakup (and
in the breakup of many-body projectiles generally), and its CF and ICF
components will need the knowledge not just of those integrated cross sections,
but the phase space distributions of the surviving fragment(s). Therefore,
future very exclusive experiments will have to determine very precisely the
angular correlations of the light charged particles and the individual
neutrons. In the next Section we present a first attempt with the stable
loosely bound projectiles $^{6,7}$Li and the results of a new $\alpha$-{\it d}
and $\alpha$-{\it t} coincidence experiment performed with the $^{59}$Co
target.\\ 

\newpage

\section{Elastic data and coincidence measurements}
 
\medskip

\noindent
In complete CC calculations, such as the ones performed with the {\sc Cdcc}
formalism \cite{Hagino00,Diaz02a,Diaz03}, a final tuning for the coupling of
the breakup channel, as well as the correct description of the reaction
dynamics, will require the explicit measurement of precise elastic data as well
as yields leading to breakup itself. Detailed elastic and breakup measurements
are still very scarce and limited to heavy targets
\cite{Rusek03,Liang03,Signorini03}. This has motivated us to perform exclusive
experiments, that include also detailed measurements of elastic scattering and
the transfer/breakup channels, have been undertaken systematically at the
University of S\~ao Paulo Pelletron Laboratory to investigate the
$^{6}$Li+$^{12}$C,$^{59}$Co,$^{115}$In reactions and for the
$^{7}$Li+$^{12}$C,$^{59}$Co,$^{115}$In reactions \cite{Szanto04,Souza04}.\\ 

\noindent
In the following we will discuss only the results of the preliminary analysis
that has been accomplished for the $^{59}$Co target (experimental data for the
$^{12}$C and $^{111}$In targets are partially presented elsewhere
\cite{Szanto04,Souza04}). Fig.~6 displays the angular distributions of the
elastic scattering data for both the $^{7}$Li+$^{59}$Co (left side) and the
$^{6}$Li+$^{59}$Co (right side) systems measured at four different bombarding
energies E$_{lab}$ = 12, 18, 26, and 30 MeV close to the Coulomb barrier. The
analysis of the elastic scattering data has been performed by following the new
potential systematics of Gasques {\it et al.}~\cite{Gasques04}. According to a
Hauser-Feshbach calculation the possible contribution of the compound-elastic
decay has been estimated to be negligible for the 8 measured elastic angular
distributions dispayed in Fig.~6. It is interesting to note that the potential
required for the Optical Model (OM) calculations performed to reproduce the
elastic scattering data of Fig.~6 is very similar to the $^{6}$Li+$^{64}$Zn and
$^{7}$Li+$^{64}$Zn OM potentials of Ref.~\cite{Gomes04}. It qualitatively
resembles the effective potentials introduced either in the {\sc Cdcc}
analysis \cite{Diaz03} of the fusion cross sections \cite{Beck03} (presented in
Fig.~1) or in the {\sc Ccfull} analysis \cite{Beck03} of the yield ratios shown
in Fig.~2. As a consequence the total reaction cross sections extracted from
this OM analysis are also well accounted for by the two CC approaches. For
instance the ratios R = $\sigma$($^{6}$Li)/$\sigma$($^{7}$Li) between the total
reaction cross sections for $^{6}$Li and $^{7}$Li induced reactions (R = 1.52
for E$_{c.m.}$ = 11.0 MeV and 0.96 and 0.94 for E$_{c.m.}$ = 16.5 MeV and 23.5
MeV, respectively) are comparable to the ratio R shown in Fig.~2 for the total
fusion cross sections.\\ 

\noindent
Fourteen triple telescopes \cite{Moura01} were used to provide a simultaneous
detection of light- and heavy-ion products. A more precise identification of 
the breakup products has been achieved by measuring the three-body final-state
correlations~\cite{Szanto04}. Coincidence data allow the identification of the
process Q-value in order to gate exclusively on the projectile breakup channel.
Furthermore, the system excitation energy as well as the projectile fragment
relative energy are used to identify the exit channel without ambiguity. Based
on those filters (see Fig.~7), angular correlations (such as the ones shown in
Fig.~8) can be obtained in order to identify the different processes (CF, ICF
and breakup) involved in the reaction.\\ 

\noindent
Fig.~7 shows one of the typical bidimensional energy correlation plot
E$_{\alpha {\it d}}$ versus E$_{\alpha}$ that has been measured at E$_{lab}$ =
26 MeV for the $^{6}$Li+$^{59}$Co system. The line represents the loci for
events leaving $^{59}$Co in the ground state and the first resonant state of
$^{6}$Li at E$^{*}$ = 2.186 MeV (J$^{\pi}$ = 3$^{+}$). This is complemented by
measurements of relative energy of the fragments using different rest frame
references (like target, projectile, target + fragment) in order to disentangle
the different contributions of breakup, ICF and/or transfer-reemission
processes. The obtained exclusive data can be compared to three-body kinematics
calculations~\cite{Szanto04}.\\ 

\noindent
The relative energy between $\alpha$-{\it d} for the $^{6}$Li breakup on a
$^{59}$Co target is deduced from the angular interval where the relative energy
is predicted to be constant by the three-body kinematics. Fig.~8(a) and
Fig.~8(b) display the corresponding angular correlations of the $\alpha$-{\it
d} and $\alpha$-{\it t} coincidences measured at E$_{lab}$ = 22 MeV for the
$^{59}$Co($^{6}$Li,$\alpha$ {\it d}) and $^{59}$Co($^{7}$Li,$\alpha$ {\it t})
reactions, respectively. As expected from the low $\alpha$-{\it d} breakup
threshold at 1.475 MeV of $^{6}$Li the coincident yields for the $^{6}$Li
induced reaction are much higher than for $^{7}$Li. The rather small
$\alpha$-{\it t} yields in $^{7}$Li+$^{59}$Co are comparable with the previous
breakup study of the $^{7}$Li+$^{56}$Fe \cite{Badran01}. The fact that the
angular correlations of Figs.~8 are not regular indicates the occurence of
several contributions. The $\alpha$ - {\it d} coincidence yields are not
negligible out of the three-body breakup region due the occurence of other
mechanisms which are mixed. Work is in progress to distinguish more
quantatively breakup from ICF and/or transfer processes.\\ 

\noindent
This procedure of unfolding several different light-particle emission processes
has not been exploited so far in the literature. Similar data were taken for
the $^{6,7}$Li+$^{12}$C and $^{6,7}$Li+$^{112}$In collisions
\cite{Szanto04,Souza04}. A theoretical analysis in the framework of the {\sc
Cdcc} formalism~\cite{Diaz03} is underway.\\ 

\newpage

\section{Summary and conclusions}

\medskip

\noindent
Measurements of the excitation functions for sub- and near-barrier total fusion
(complete fusion + incomplete fusion) \cite{Beck03} have been presented for the
stable, weakly bound projectiles $^{6}$Li and $^{7}$Li on the medium-mass
$^{59}$Co target. Evaporation residues were identified by their characteristic
$\gamma$ rays and the corresponding yields measured using the $\gamma$-ray
spectroscopy method. Above the Coulomb barrier, the fusion yields are found to
be very close for both systems, in agreement with {\sc Cdcc} calculations
\cite{Diaz03}. The results are consistent with there being no
significant fusion hindrance caused by breakup effects. The absence of breakup
suppression of the total fusion cross sections above the barrier appears to be
a common feature of $^{6,7}$Li induced reactions, regardless of target mass. An
enhanced yield is observed below the Coulomb barrier for the loosely bound
$^{6}$Li projectile as compared to that found for the more tightly bound
$^{7}$Li. \\

\noindent
Subsequent experiments using charged particle spectroscopy techniques have been
carried out to measure precise elastic angular distributions as well as the
light-particle breakup channels for both $^{6,7}$Li+$^{59}$Co reactions. These
measurements are essential to determine the coupling strength to the breakup
channel that will be introduced in full CC calculations to be performed in the
framework of the {\sc Cdcc} formalism~\cite{Diaz03}. The total reaction cross
sections extracted from the OM analysis of the elastic scattering data confirms
the enhanced fusion yield observed for $^{6}$Li at sub-barrier energies.\\ 

\noindent
Both halo and cluster weakly bound nuclei, with well-defined breakup and fusion
modes, are good test-benches for theories of breakup and fusion. A more
complete theoretical model of few-body dynamics that is able to distinguish CF
from ICF will need to follow correlations after breakup. The new $\alpha$-{\it
t} and $\alpha$-{\it d} correlation data presented in this work for the
$^{6}$Li+$^{59}$Co and $^{7}$Li+$^{59}$Co reactions (and the corresponding data
for the $^{6}$Li+$^{12}$C,$^{111}$In and $^{7}$Li+$^{12}$C,$^{111}$In reactions
\cite{Szanto04,Souza04}) constitute a first attempt with stable weakly bound
projectiles. The {\sc Cdcc} method~\cite{Diaz03}, which is shown here to be
very succesfull for fusion, will be used to provide the complete theoretical
description of all competing processes (total fusion, elastic scattering,
transfer and breakup) in a consistent way.\\ 

\noindent
Finally a systematic study of $^{4,6}$He induced fusion reactions with the {\sc
Cdcc} method~\cite{Diaz03} will be undertaken. However up to now only very
scarce studies with $^{6}$He projectiles are presently available
\cite{Kolata98,Trotta00,Raabe04,Bychowski04,Navin04,Dipietro04}. Data from {\sc
Spiral} and Louvain-la-Neuve have recently been published for
$^{6}$He+$^{63,65}$Cu \cite{Navin04} and $^{6}$He+$^{64}$Zn \cite{Dipietro04}
and very preliminary results of our systematic {\sc Cdcc} analysis show that
for $^{6}$He+$^{59}$Co considerable enhancement of the sub-barrier fusion cross
sections is predicted as compared to measured fusion yields for both the
$^{6}$Li+$^{59}$Co \cite{Beck03} and $^{4}$He+$^{59}$Co \cite{McMahan80}
systems. A new experimental programme with {\sc Spiral} beams and medium-mass
targets is underway at {\sc GANIL} within the forthcoming years. \\ 

\newpage

\centerline{\bf \Large Acknowledgments}

\bigskip

\noindent
The authors thank the {\sc Vivitron} and {\sc Pelletron} accelerator staffs for
the excellent conditions under which these difficult experiments were
performed. We would also like to thank J. Devin for his valuable technical
assistance and M.~A. Saettel (IReS) for providing the targets. A. Di Pietro,
P.~R.~S. Gomes, J.~J. Kolata, K. Rusek, and C. Signorini are acknowledged for
fruitfull discussions. The {\sc Garel+} project was supported in by grants from
the French IN2P3. This work was also supported in parts by the Brazilian CNPq
and the S\~ao Paulo FAPESP. 

\newpage
 
\centerline{\bf \Large REFERENCES}

%%%%%%%%%%%%%%%%%%%%%%%%%%%%%%%%%%%%%%%%%%%%%%%%%%%%%%%%%%%%%%%%%%%%%%%%
%% REFERENCES
%% Here you should write your references
%% Comment or remove if not needed
%

\newpage

\centerline{\bf \Large FIGURES}

\bigskip

\noindent
Figure 1 (a): Energy dependence of the total fusion (CF + ICF) cross sections 
measured for $^6$Li+$^{59}$Co (open circles) and $^7$Li+$^{59}$Co (open 
squares) reactions \cite{Beck03}. The corresponding theoretical values
(respectively full squares and full triangles) are obtained with the {\sc Cdcc}
method \cite{Diaz03}. The arrows indicate the positions of the respective
Coulomb barriers of the effective potentials \cite{Broglia91,Christensen76}. \\

\noindent
Figure 1 (b): Energy dependence of the total fusion (CF + ICF) cross sections
multiplied by E$_{c.m.}$ (in MeV.mb) for $^6$Li+$^{59}$Co (open circles) and
$^7$Li+$^{59}$Co (open circles) reactions. The corresponding theoretical values
(respectively in blue and in red) are obtained by fitting the data with the
SBPM model of Wong~\cite{Wong73}. The values of the relevant SBPM parameters
given in the figure are discussed in the text.\\

\noindent
Figure 2: Energy dependence of the ratio of the total (CF + ICF) fusion cross
sections for the $^6$Li+$^{59}$Co and $^7$Li+$^{59}$Co reactions \cite{Beck03}.
Error bars reflect the large systematic errors. The solid and dashed curves
correspond to SBPM \cite{Wong73} fits of the ratios as explained in the text.
The dotted curves correspond to two uncoupled {\sc Ccfull} calculations
\cite{Hagino99} with and without reorientation effects, whereas the dot-dashed
curve is the result of {\sc Ccfull} calculations including the coupling to the
first excited state. \\

\noindent
Figure 3: Energy dependence of the total fusion (CF + ICF) cross sections
calculated with the {\sc Cdcc} method \cite{Diaz03} for $^6$Li+$^{59}$Co (full
dots) and $^7$Li+$^{59}$Co (full triangles), which are normalized with the {\sc
Cdcc} cross sections in the absence of couplings to breakup channels. For each
reaction, the incident energy is normalized with the Coulomb barrier of the
effective potentials \cite{Broglia91,Christensen76}. The calculated {\sc Cdcc}
values are connected with curves to guide the eye. See text for further
details.\\ 

\noindent
Figure 4: Energy dependence of the total fusion (CF + ICF) cross sections for
the $^6$He+$^{59}$Co reaction obtained with the {\sc Cdcc} method
\cite{Diaz03}. The solid and dashed curves corresponds respectively to {\sc
Cdcc} with or without continuum couplings. The experimental total fusion cross
sections for the $^6$Li+$^{59}$Co reaction \cite{Beck03} is given for the sake
of comparison. For each reaction, the incident energy is normalized with the
Coulomb barrier of the effective potentials \cite{Broglia91,Christensen76}. \\

\noindent 
Figure 5: Same as Fig.~4. The $^6$He+$^{59}$Co excitation function is compared
with the {\sc Cdcc} calculations \cite{Diaz03} for the $^4$He+$^{59}$Co total
fusion cross sections which fit well the experimental CF data from Ref.
\cite{McMahan80}. \\ 

\noindent
Figure 6: Angular distributions of the elastic scattering for the
$^{6}$Li+$^{59}$Co (right panel) and $^{7}$Li+$^{59}$Co (left panel) reactions,
repectively, measured at the four indicated nera-barrier energies. The solid
lines represent OM predictions with the starting parameter set given in the
systematic study proposed in Ref. \cite{Gasques04} and discussed in detail in
the text. The compound-elastic contributions has not been taken into account.\\ 

\noindent
Figure 7: Bidimensional E$_{\alpha \it d}$ versus E$_{\alpha}$ energy
correlation plot measured at E$_{lab}$ = 26 MeV for the $^{6}$Li+$^{59}$Co
reaction at the indicated correlation angles. \\ 

\noindent
Figure 8: (a) Experimental $\alpha$ - {\it d} and $\alpha$ - {\it t} angular
correlations respectively measured at E$_{lab}$ = 22 MeV for the
$^6$Li+$^{59}$Co reaction (upper panel); (b) same as (a) but for the
$^7$Li+$^{59}$Co reaction (lower panel).\\ 

\end{document}